\documentclass[12pt]{article}

\usepackage[utf8]{inputenc}
\usepackage[T1]{fontenc}
\usepackage{booktabs}           
\usepackage{array}              
\usepackage{threeparttable}     
\usepackage{multirow}           
\usepackage{graphicx}           
\usepackage{amsmath}            
\usepackage{natbib}             
\usepackage{url}                
\usepackage[margin=1in]{geometry} 
\usepackage{setspace}           
\usepackage{caption}            
\usepackage{subcaption}         
\usepackage{float}              

\title{Unintended Consequences of Early Driving Access: Evidence from Graduated Driver Licensing Policies and Adolescent Health Outcomes}

\author{Sharareh Massahi\\
Department of Economics, University of Oklahoma, Norman, OK 73019, USA\thanks{Email: \texttt{sharareh.massahi@ou.edu}}}

\date{} 

\begin{document}

\maketitle
\begin{abstract}
Graduated driver licensing systems effectively reduce adolescent traffic fatalities but create unintended health consequences. Using state-level variation in licensing policies from 1999-2020 and difference-in-differences analysis, we provide the first causal evidence that early driving access generates significant health risks for female adolescents aged 15-19. States allowing learner's permits before age 16 experienced sharp increases in drug-related mortality (+1.331 per 100,000, $p<0.001$) and mental health-related mortality (+0.760, $p<0.001$), even as vehicle deaths declined (-0.656, $p<0.05$). These effects explain nearly one-third of rising adolescent drug mortality and one-tenth of mental health mortality increases over the study period.

Early driving access expands geographic reach, enabling contact with illicit drug markets previously inaccessible to adolescents. It broadens social networks, increasing exposure to high-risk peers, while vehicles provide unsupervised spaces for experimentation. Premature independence also intensifies psychological stress during critical developmental stages.

Nationally, results correspond to approximately 138 additional drug deaths and 79 mental health deaths annually among female adolescents, imposing over \$2 billion in mortality costs yearly. These findings fundamentally reshape cost-benefit assessments of licensing policies, revealing how interventions protecting adolescents in one domain can create risks in others. Evidence-based modifications including enhanced supervision, geographic restrictions, and mental health integration could preserve traffic safety gains while mitigating unintended harms. This research demonstrates the critical need for multi-outcome policy evaluation that captures both benefits and hidden costs of expanded adolescent independence.
\end{abstract}

\textbf{Keywords}: Graduated Driver Licensing, Vehicle Mortality, Drug Mortality, Mental Health, Difference-in-Differences

\textbf{JEL Classification}: I10, I12, I18, R41

\clearpage


\section{Introduction}

Across many societies, the transition to independent driving is regarded as a defining moment in adolescence, symbolizing freedom, responsibility, and the move toward adulthood. While this step is often celebrated as a positive marker of growth, it can also reshape patterns of health and safety in ways that extend far beyond traffic outcomes. The question of how early driving access influences adolescent well-being has therefore become an important area of inquiry for public health and policy.

In the United States, where our study is based, the evidence shows that adolescent mortality patterns have shifted in striking and concerning directions over the past two decades. Between 2019 and 2021, overdose deaths among youth aged 10 to 19 more than doubled, despite overall drug use among teenagers showing signs of decline \citep{tanz2022drug}. In 2021, an average of 68 adolescents lost their lives each month to drug overdoses, with illicitly manufactured fentanyls responsible for the overwhelming majority of cases. These losses have fallen disproportionately on teenage girls, a group once thought less vulnerable to the most severe consequences of drug epidemics.

The environment confronting today’s adolescents differs substantially from that of previous generations. \citet{case2015rising} described how “deaths of despair” among middle-aged Americans were driven by economic insecurity, erosion of social connections, and the growing availability of lethal substances. Although their work focused on adults, the same structural forces increasingly affect younger populations. Research has demonstrated the complex bidirectional relationships between health outcomes, economic inequality, and growth, showing how income disparities can negatively affect health indicators while improved health outcomes contribute to broader societal development \citep{hosseinidoust2021investigating}. As adolescents gain independence through expanded mobility, they encounter greater exposure to the very conditions that have fueled broader mortality crises, including access to dangerous substances, high-risk peer environments, and psychological pressures linked to premature autonomy.

This study contributes to understanding these dynamics by examining the unintended health consequences of graduated driver licensing (GDL) systems. While such policies were designed to reduce traffic fatalities and have been successful in doing so, our analysis shows that early driving access also creates measurable increases in drug-related and mental health mortality among female adolescents. The relationship between health outcomes and policy interventions is complex, and research has demonstrated that health-related expenditures and interventions can have significant positive effects on broader social and economic indicators, including life expectancy and overall population well-being \citep{massahi2021effect}. These findings underscore the importance of evaluating policies across multiple outcome domains, rather than focusing narrowly on their intended effects.

\section{Background and Conceptual Framework}

Understanding the consequences of driving access for adolescents requires more than a narrow focus on road safety. Driving is not only a means of transportation but also a powerful gateway to autonomy, shaping social interactions, daily routines, and access to both protective and harmful environments. For this reason, evaluating policies such as GDL demands a broader lens that considers multiple domains of adolescent health and well-being.

\subsection{The Evolution of Risk in the Digital Age}

The landscape of drug use among young people has changed dramatically in recent years, creating risks that earlier generations did not face. Counterfeit pills containing potent and often deadly substances are now widely available, frequently designed to appear indistinguishable from legitimate medications \citep{tanz2022drug}. For adolescents, who may perceive prescription drugs as relatively safe, this shift represents a dangerous recalibration of risk. Substances that once carried relatively low lethality can now result in immediate and irreversible harm.

The channels through which teenagers gain access to these substances have also transformed. Social media and online platforms have increasingly replaced older, more visible pathways of distribution, allowing adolescents to obtain drugs with unprecedented ease \citep{singh2019opioid}. Unlike traditional networks, which often involve face-to-face interactions and some level of adult oversight, these digital spaces offer few opportunities for intervention. For teenage girls, who consistently report higher levels of engagement with social media compared to their male peers, this shift is particularly alarming. Greater exposure to online environments not only increases opportunities for contact with illicit sellers but also normalizes risky behaviors within peer groups.

Equally troubling are the changes in adolescent mental health that have unfolded alongside the rise of lethal drug markets. Evidence shows a sharp rise in intentional self-poisonings, particularly among young women. Between 1998 and 2014, intentional poisoning cases increased by more than 50 percent among individuals aged 10 to 24, with the steepest increases observed in girls aged 16 to 18 \citep{tyrrell2017differing}. These findings mirror broader trends in adolescent psychological distress: higher reported rates of depression, anxiety, and suicidal ideation, conditions strongly linked to substance misuse and overdose risk \citep{nosrati2019economic}. The convergence of declining mental health and the ready availability of lethal substances has created a uniquely dangerous environment for adolescents, especially for girls who face a disproportionate share of these challenges.

\subsection{Driving Access as a Policy Intervention with Unintended Consequences}

GDL systems were introduced across all U.S. states with a clear purpose: to reduce traffic accidents among young drivers. By phasing in driving privileges through stages of supervised and restricted licenses, these programs succeeded in lowering vehicle-related fatalities and remain one of the most significant public health achievements in traffic safety. Yet the same policies that improved road outcomes also created differences in the age at which adolescents gain independent mobility, offering a unique setting to study broader consequences of early driving access.

Recent quasi-experimental evidence underscores the importance of this perspective. \citet{huh2021teenage} showed that receiving a driver’s license can fundamentally change adolescent behavior, increasing risky activities and raising the likelihood of substance-related harms. Their findings revealed that poisoning deaths, primarily overdoses, rose by nearly 78 percent among teenage females after gaining independent driving privileges. Such results highlight that driving does not merely grant mobility; it reshapes the opportunities and environments to which adolescents are exposed.

The pathways through which driving access intensifies risk are multiple and interconnected. First, mobility extends the geographic reach of adolescents, enabling them to travel to areas with greater drug availability and connect with markets otherwise out of reach. Second, expanded travel increases the size and diversity of peer networks, which can include substance-involved groups beyond the immediate neighborhood or school environment. Third, vehicles themselves become private spaces, creating opportunities for experimentation without the watchful eye of parents or guardians. Finally, driving provides a logistical advantage, allowing teenagers to coordinate purchases, transport substances, and engage in complex behaviors that would be difficult without independent transportation.

Taken together, these mechanisms demonstrate that policies intended to enhance safety in one domain can unintentionally shift risks into another. Early driving access reduces exposure to traffic dangers through structured licensing, but it also accelerates exposure to drug markets, risky peer groups, and unsupervised environments. For female adolescents, who already face heightened vulnerability to both substance use and mental health challenges, this trade-off is especially consequential.

\subsection{Structural Vulnerabilities and Policy Interactions}

The risks facing adolescents, and particularly teenage girls, cannot be understood only through individual choices. Broader structural conditions shape both exposure and vulnerability. Economic decline, social dislocation, and mass incarceration have been shown to drive increases in drug-related mortality across U.S. counties \citep{nosrati2019economic}. When teenagers gain independence through driving, they also become more exposed to these systemic challenges. For young people already navigating untreated mental health difficulties, limited coping resources, and constrained access to support systems, the additional autonomy of driving can heighten rather than reduce risk.

Policymakers have sought to respond with interventions such as Prescription Drug Monitoring Programs (PDMPs). Evidence indicates that mandatory PDMPs reduce prescription drug misuse and opioid-related deaths among young adults \citep{grecu2019mandatory, meinhofer2018prescription}. Yet the effectiveness of these programs has weakened as drug markets have shifted toward illicit fentanyl and counterfeit pills that operate outside medical systems. In this new environment, the enhanced mobility that driving provides can unintentionally facilitate access to substances that lie beyond the reach of formal regulation. The interaction of structural disadvantage, fragile mental health, and increased autonomy creates what can be described as a “perfect storm” of vulnerability for teenage girls. Independence that was intended to foster resilience can instead magnify exposure to danger. 

\subsection{Critical Gaps in Current Understanding}

Despite mounting evidence of these risks, important gaps remain in the literature. First, most studies rely on observational data that cannot establish causal relationships between driving access and health outcomes. Although \citet{huh2021teenage} offered quasi-experimental evidence, their work did not fully explore gender differences or disaggregate specific mortality categories. Second, research often treats driving, mental health, and substance use as separate issues rather than examining how they interact to produce compounded risks. Third, evaluations of GDL policies tend to focus narrowly on traffic safety, overlooking how these policies intersect with drug markets and adolescent mental health. Finally, most prior studies concentrate on single outcomes, limiting our ability to understand the broader spillover effects that are essential for accurate cost-benefit assessments.

\subsection{Study Objectives and Contribution}

This study addresses these gaps by providing the first comprehensive causal analysis of how GDL policies influence non-traffic health outcomes for female adolescents. Using state-level variation in minimum driving ages from 1999 to 2020, we employ a difference-in-differences (DiD) approach to examine three key categories of mortality: drug-related deaths, mental health-related deaths, and vehicle fatalities. This design allows us to identify the broader consequences of adolescent mobility while addressing confounding factors that have limited earlier research.

By focusing on female adolescents, a group showing heightened vulnerability to both psychological distress and substance-related harm, the study offers targeted evidence with direct policy relevance. Methodologically, our approach treats state-level licensing reforms as a natural experiment, enabling a credible assessment of causal effects. Substantively, we move beyond single-outcome analyses by considering multiple domains of mortality, highlighting the trade-offs that arise when policies designed to solve one problem inadvertently create others. 

The findings contribute to both transportation safety and public health research by showing that early driving access reduces vehicle deaths while simultaneously increasing drug- and mental-health-related mortality. Recognizing this duality is critical for policymakers. Evidence-based modifications to GDL systems—such as stronger supervision requirements, geographic restrictions, and integration with mental health support—can help preserve the gains in traffic safety while mitigating the unintended costs. More broadly, this study demonstrates the importance of comprehensive evaluation frameworks that account for multiple outcomes when assessing policies that reshape adolescent independence.


\section{Methodology}

\subsection{Research Design}

To investigate how early driving access affects adolescent health, we adopt a DiD design that exploits state-level variation in the timing and structure of GDL policies. Our analysis focuses on female adolescents aged 15 to 19 years in the United States over the period 1999 to 2020. This quasi-experimental framework leverages the staggered rollout of GDL systems and variation in minimum driving ages across states to identify the causal effects of adolescent mobility on mortality outcomes.

The DiD approach is particularly well-suited to this setting because it addresses several key identification challenges. First, states differ in cultural, economic, and demographic characteristics that also shape both mortality risks and the adoption of licensing policies. By including state fixed effects, we account for all such time-invariant differences. Second, nationwide shocks—such as economic downturns, public health campaigns, or shifts in drug markets—could simultaneously affect mortality outcomes in all states. Year fixed effects capture these common temporal influences. Third, the staggered nature of GDL adoption across states provides meaningful variation in treatment timing, allowing us to strengthen causal inference by comparing states before and after their own reforms as well as relative to others.

Central to the credibility of this strategy is the parallel trends assumption, which requires that mortality trajectories in treatment and control states would have followed similar paths in the absence of policy changes. We assess this assumption through a combination of visual inspection of pre-treatment mortality patterns, formal pre-trend tests, and event study analyses. In addition, placebo tests with randomly assigned treatment timing help verify that our results are not driven by spurious correlations. Together, these approaches provide strong support for the validity of our empirical framework.

\subsection{Data Sources and Sample Construction}

Our dataset brings together multiple authoritative sources to create a comprehensive state-year panel covering all 50 states and the District of Columbia from 1999 to 2020. This twenty-two–year observation window spans the full period of GDL adoption while providing sufficient pre- and post-treatment years to evaluate policy impacts. The resulting panel enables us to study adolescent mortality outcomes with both breadth and depth, capturing long-term patterns while maintaining granular state-level detail.

Mortality outcomes are drawn from the CDC Multiple Cause of Death files \citep{cdc2023}, which represent the most complete enumeration of deaths in the United States. These data are rigorously compiled by the National Center for Health Statistics and are widely regarded as the gold standard for mortality research. We extract deaths among females aged 15–19 years and classify them by underlying cause using International Classification of Diseases, Tenth Revision (ICD-10) codes. This coding system allows precise differentiation between drug-related, mental health-related, and vehicle-related mortality, which form the core outcomes of our analysis.

Building on these data, we define three distinct categories of mortality outcomes. Each category is carefully constructed to capture both the primary pathways through which adolescent health is affected and the broader consequences of mobility access.

Mental health-related mortality is defined to include all deaths coded as intentional self-harm or suicide (ICD-10 codes X60–X84). To capture the broader role of mental health conditions, we also incorporate deaths where psychiatric disorders were documented in the multiple cause fields. This inclusive approach ensures that we measure not only direct suicides but also fatalities where untreated or undiagnosed mental health struggles contributed to the outcome. The resulting measure provides a more accurate reflection of the mental health crisis among adolescents than suicide statistics alone.

Drug-related mortality encompasses all deaths attributed to drug poisoning, regardless of intent. This includes accidental overdoses (X40–X44), intentional self-poisonings (X60–X64), assaults using drugs (X85), and poisonings of undetermined intent (Y10–Y14). We adopt this broad definition to acknowledge the difficulty of distinguishing intent in adolescent populations, where circumstances often remain ambiguous. Within this category, we pay special attention to opioid-related codes, which allow us to track the disproportionate impact of the opioid epidemic on adolescent health.

Vehicle-related mortality covers all transport accidents (ICD-10 codes V01–V99), including fatalities among drivers, passengers, pedestrians, and cyclists. This category directly reflects the original intent of GDL policies: improving road safety. By including this measure alongside drug- and mental health-related mortality, we are able to test both the expected benefits of early driving restrictions and the unintended consequences into other areas of adolescent risk.

To complement the mortality data, we compile a detailed dataset on state driver licensing policies. Using information from the Insurance Institute for Highway Safety \citep{iihs2023} and individual state Departments of Motor Vehicles \citep{StateDMV2020}, we systematically review statutory language and policy changes across all states. This allows us to document the exact timing of GDL implementation, the minimum ages for learner’s permits and unrestricted licenses, and key provisions such as passenger limits, nighttime driving restrictions, and required supervision. These institutional details form the foundation of our treatment classification, ensuring that our analysis reflects meaningful variation in adolescent mobility access.

Our treatment classification focuses on states that allow adolescents to obtain a learner’s permit before the age of sixteen, which we define as \textit{early driving access}. These states provide an earlier pathway to independent mobility compared to those that maintain the standard age-16 threshold. The treatment group includes Alabama, Alaska, Arkansas, Colorado, Florida, and Georgia. Control states, by contrast, require adolescents to wait until at least age 16 for a learner’s permit; examples include Arizona, California, Connecticut, and Delaware. This grouping captures meaningful variation in the timing of adolescent mobility while maintaining sufficient sample size for robust statistical inference. 

To strengthen causal inference, we incorporate a comprehensive set of state-level control variables that account for differences in economic conditions, healthcare access, policy environments, and geography. These controls ensure that our estimated effects reflect the consequences of early driving access rather than confounding influences.

Economic conditions are measured using indicators such as unemployment rates, median household income, and poverty rates. These data are drawn from the U.S. Census Bureau’s Small Area Income and Poverty Estimates \citep{census2023} and the Bureau of Labor Statistics \citep{bls2023}. Controlling for economic context is critical because both adolescent health outcomes and political willingness to adopt policy reforms often vary with macroeconomic trends.

Healthcare access variables are derived from the Area Health Resources Files maintained by the Health Resources and Services Administration \citep{hrsa2023}. We include the number of primary care physicians and mental health providers per 100,000 population. These measures account for the local healthcare infrastructure available to adolescents, recognizing that access to treatment and prevention services may influence mortality outcomes independently of mobility policies.

Policy environment controls capture other state-level interventions that could affect adolescent mortality. We include indicators for active Prescription Drug Monitoring Programs (PDMPs) and for mandatory prescriber consultation requirements, using data from the PDMP Training and Technical Assistance Center \citep{pdmp2023}. In addition, we measure the availability of substance abuse treatment facilities per 100,000 residents, using the National Directory of Drug and Alcohol Abuse Treatment Programs published by the Substance Abuse and Mental Health Services Administration \citep{samhsa1999-2020}. These controls ensure that the estimated impact of early driving access is not conflated with contemporaneous changes in drug regulation or treatment capacity.

Geographic factors are captured through regional indicators and rural–urban classification codes from the U.S. Department of Agriculture Economic Research Service \citep{usda2023}. These measures control for systematic differences in both mortality and policy adoption patterns across geographic settings. Rural states, for example, often face higher driving exposure but limited access to healthcare facilities, whereas urban states may exhibit opposite dynamics. Including these variables helps isolate the specific role of driving access in shaping adolescent health outcomes.

\subsection{Empirical Strategy}

Our primary empirical specification is a DiD model of the following form:

\begin{equation}
MR_{ist} = \alpha + \beta(EarlyDriving_i \times PostGDL_t) + \gamma X_{ist} + \delta_s + \theta_t + \varepsilon_{ist}
\end{equation}

Where $MR_{ist}$ is the mortality rate per 100,000 female adolescents aged 15–19 in state $s$ and year $t$. The key parameter of interest, $\beta$, measures the effect of early driving access on adolescent mortality. $EarlyDriving_i$ is an indicator for states that allow permits before age 16, while $PostGDL_t$ denotes the period after widespread adoption of GDL systems(beginning in 2000). The interaction of these terms captures the differential impact of early access after GDL implementation. $X_{ist}$ represents the vector of state-level control variables described above, while $\delta_s$ and $\theta_t$ are state and year fixed effects, respectively. Standard errors are clustered at the state level to account for serial correlation within states over time.

The credibility of this approach rests on two main identification assumptions. First, the parallel trends assumption requires that treatment and control states would have followed similar mortality trajectories in the absence of policy changes. We evaluate this assumption using visual inspection of pre-treatment mortality patterns, statistical tests of pre-trends, and event study analyses. Second, the exogeneity assumption requires that the timing of GDL adoption and minimum driving age policies is not endogenously determined by adolescent mortality. Historical evidence suggests that adoption was driven primarily by traffic safety concerns and federal pressure in the late 1990s, rather than by local health crises, which supports this assumption.

To further validate our identification strategy, we conduct a series of robustness checks. Event study models trace the evolution of treatment effects before and after policy implementation, allowing us to test for anticipatory effects and to examine the dynamics of policy impact over time. Placebo tests, in which treatment status is randomly assigned to non-treatment states, confirm that our estimates are not driven by spurious correlations. We also assess robustness to alternative clustering strategies, different sets of fixed effects, and varying definitions of the post-treatment period.

\subsection{Statistical Analysis and Robustness}

All statistical analyses are performed in Stata/SE 17.0. Significance is evaluated at conventional thresholds (1\%, 5\%, and 10\%). Standard errors are clustered at the state level to correct for within-state correlation across years, which is crucial in a panel setting with policy variation at the state level.

Our robustness exercises include alternative clustering (e.g., regional clustering), sensitivity checks using robust standard errors, and sample restrictions that exclude early adopters or focus on specific time windows. These tests ensure that our results are not dependent on particular modeling choices or outlier states. Diagnostic tests—such as pre-treatment trend analysis and placebo interventions—reinforce the credibility of our findings. 

Finally, by estimating separate models for drug-related, mental health-related, and vehicle-related mortality, we are able to assess whether observed effects are concentrated in specific outcome domains. This multi-outcome framework not only strengthens the validity of our conclusions but also highlights the trade-offs policymakers face when evaluating interventions that shape adolescent independence. Together, these methodological choices provide a rigorous and transparent foundation for the results that follow.

\section{Results}

\subsection{Main Findings}

This study provides causal evidence that policies regulating teenage driving through GDL shape adolescent health beyond traffic outcomes. Using state-level differences in minimum driving age requirements from 1999 to 2020, we find that early access to independent mobility affects three core mortality domains among female adolescents aged 15–19.

States that granted learner’s permits before age 16 experienced a sharp rise in drug-related mortality \((+1.33\ \text{per }100{,}000,\ p<0.001)\) and a parallel increase in mental health-related mortality \((+0.76\ \text{per }100{,}000,\ p<0.001)\). At the same time, these policies achieved the expected traffic safety gains, with vehicle-related mortality declining \(({-}0.66\ \text{per }100{,}000,\ p<0.05)\). Quantitatively, the policy explains approximately 29.6\% of the observed increase in drug mortality and 9.9\% of the increase in mental health mortality among teenage girls over the study period. These effects indicate that early mobility produces meaningful shifts in adolescent risk exposure, improving road safety while intensifying hazards linked to drugs and mental health.

\subsection{Treatment and Control Group Definition}

To isolate these effects, we classify states based on their minimum age requirements for learner’s permits. The \textit{treatment group} consists of states that allowed permits before age 16, thereby granting adolescents earlier independence and mobility access. This group includes Alabama, Alaska, Arkansas, Colorado, Florida, and Georgia. In contrast, the \textit{control group} consists of states that maintained the standard age-16 threshold for learner’s permits, delaying independent access to driving. This group includes all the other 45 states. By structuring the analysis in this way, we capture meaningful policy variation while ensuring sufficient heterogeneity for external validity. Importantly, the treatment states span both southern and western regions, while the control states are drawn from the Southwest, West Coast, and Northeast. This geographic and socioeconomic diversity reduces the likelihood that our estimates merely reflect regional idiosyncrasies rather than the true consequences of early driving access.

This classification captures meaningful policy variation while preserving sufficient heterogeneity for external validity. Treatment states span multiple regions of the United States, including the South and the West, and control states cover the Southwest, West Coast, and Northeast. The geographic and socioeconomic diversity on both sides reduces the risk that estimates reflect regional idiosyncrasies rather than the consequences of early driving access itself.

\subsection{Temporal Trends in Female Adolescent Mortality}

Understanding how mortality patterns evolved over the study period is essential for placing the causal effects of driving access into context. To achieve this, we examined both period-averaged and annual mortality rates among female adolescents aged 15 to 19 from 1999 to 2020. These descriptive analyses allow us to distinguish short-term fluctuations from sustained shifts and to identify the inflection points where health outcomes changed most dramatically.

\subsubsection{Annual Mortality Trends and Critical Inflection Points}

\begin{figure}[htbp]
\centering
\includegraphics[width=0.9\textwidth]{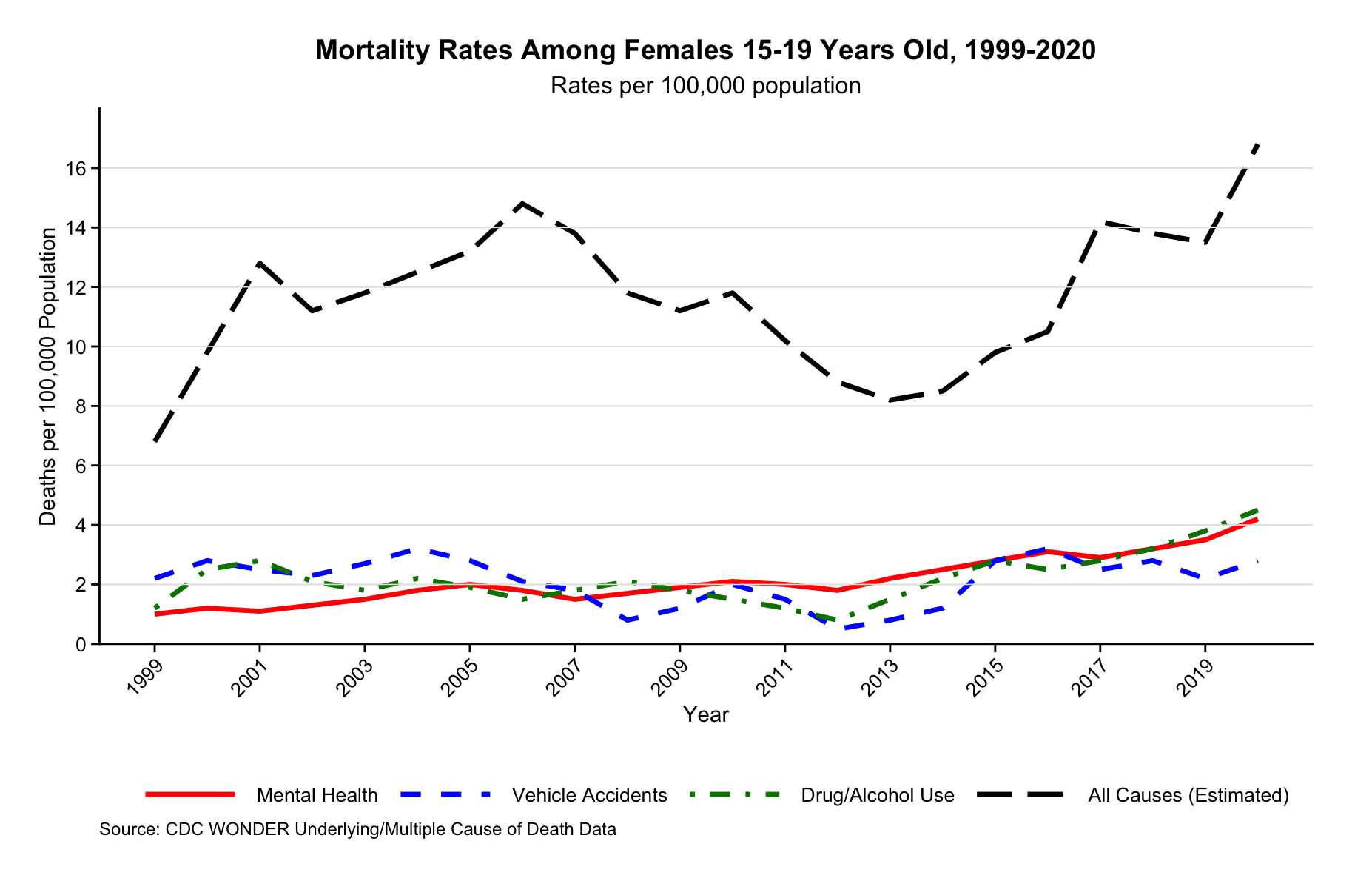}
\caption{Annual Mortality Rates Among Females 15--19 Years Old, 1999--2020}
\label{fig:annual_mortality}
\end{figure}

Figure \ref{fig:annual_mortality} shows four series for females aged 15–19: all causes, vehicle accidents, drug and alcohol, and mental health. The black all–cause line is not a smooth trend. It moves through three clear regimes. First, a volatile early phase up to the mid–2000s with two visible crests around 2000 and 2006. Second, a trough from roughly 2009 to 2013 when all–cause deaths reached their lowest levels. Third, a sustained climb after 2014 that ends at the highest point in 2020. Calling the middle phase “stable” would be misleading because the series is not flat across the entire window. It bottoms out and then pivots upward. The timing of that pivot is the critical fact in this figure.

The cause–specific lines explain why the black line bends the way it does. The blue vehicle series falls steadily through the 2000s and approaches a floor around 2011–2013. That decline is large enough to pull the black line down to its trough. After 2013, the blue line inches back up, but it never returns to its early levels. The green drug and alcohol series follows a U–shape. It is elevated in the mid–2000s, falls to a low point near 2012, then accelerates sharply from 2014 onward and finishes as the steepest line in the figure. The red mental health series climbs gradually across the entire period. It does not spike, but its slope is persistently positive, and the rise becomes steeper after 2016. By the end of the sample, the green and red lines sit close together, which matches the narrative that drug exposure and psychological distress intensified in the late 2010s.

The key contribution of this figure is compositional. The early improvement in all–cause mortality is almost entirely a vehicle story. The latter deterioration is almost entirely a drug and mental health story. In other words, the figure documents a handoff in what drives adolescent mortality. Before 2010, safer roads dominate the aggregate. After 2014, the expansion of lethal substances and worsening mental health dominated the aggregate. This handoff is visible in the curvature of the black line: a downward pull from blue in the 2000s, followed by an upward pull from green and red in the late 2010s.

Two additional details make the figure policy relevant. First, there is a lead–lag pattern. The large reduction in vehicle deaths is realized before the surge in drug and mental health deaths. That timing is consistent with early mobility reducing crash risk at first while gradually expanding access to risk environments that matter for overdose and self–harm. Second, the green and red lines moved together after 2014. Their joint rise is not a single–year shock. It is a multi–year co–movement that aligns with the period when synthetic opioids became pervasive and when adolescent mental health measures worsened. The figure, therefore, does more than describe levels. It isolates when the aggregate risk composition changed and points to why it changed, which is the empirical foundation for our difference–in–differences results that follow.

\begin{figure}[htbp]
\centering
\includegraphics[width=0.9\textwidth]{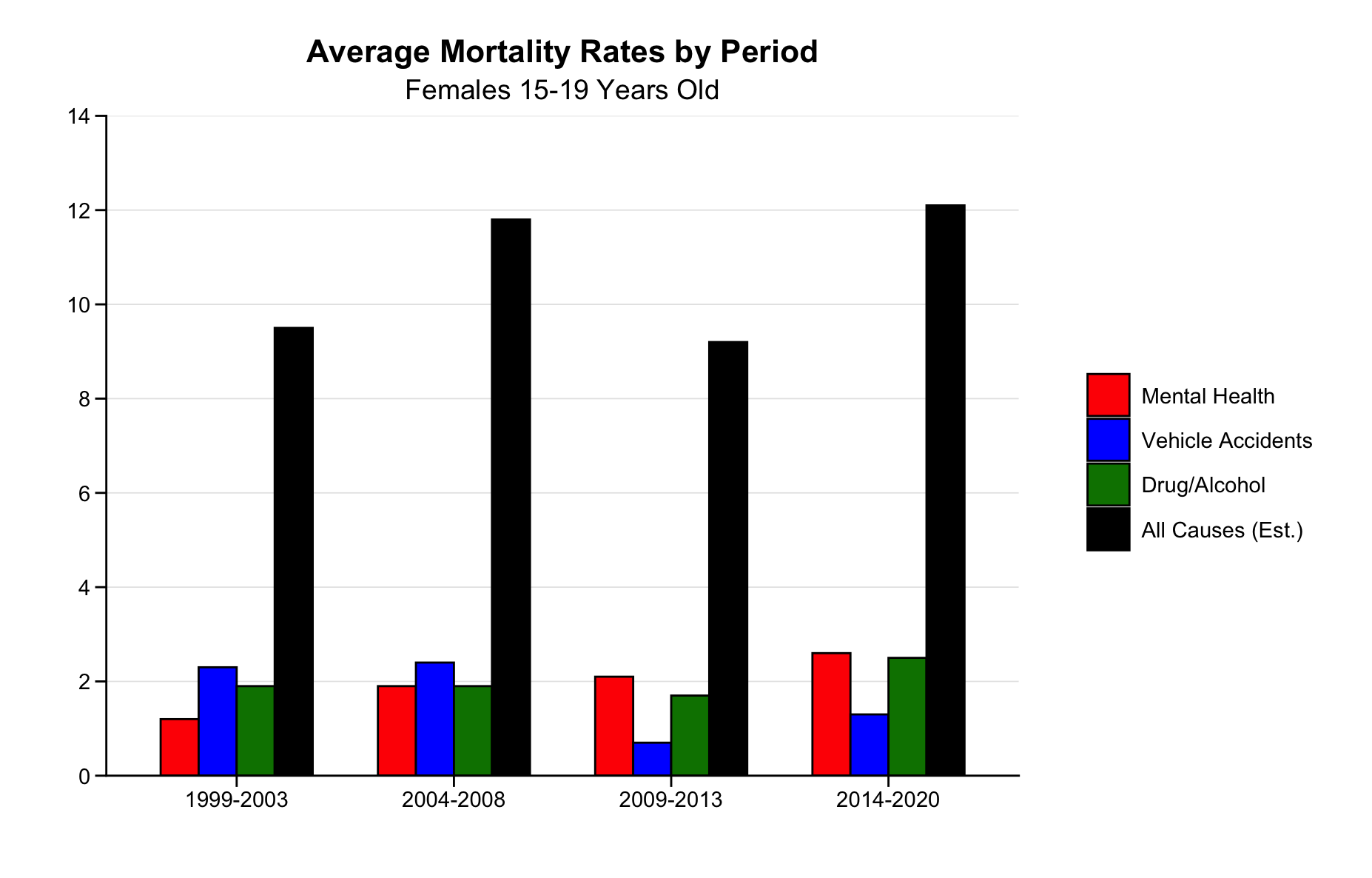}
\caption{Annual Mortality Rates by Cause, 1999--2020 -- Females 15--19 Years Old}
\label{fig:annual_by_cause}
\end{figure}

Disaggregating the annual data by cause, as shown in Figure \ref{fig:annual_by_cause}, deepens this story. Mental health mortality rose steadily throughout the entire period, quadrupling from roughly 1 to nearly 4 deaths per 100,000 between 1999 and 2020. Drug- and alcohol-related mortality remained stable until 2013, but then surged sharply, reaching levels comparable to mental health mortality by 2020. Vehicle-related mortality followed the opposite trajectory: it declined rapidly through the late 2000s, then leveled off, and modestly increased again after 2013.

Taken together, these trends identify 2014 as a pivotal inflection point in adolescent health outcomes. The simultaneous rise in drug and mental health deaths, coupled with the reversal of progress in overall mortality, signals the beginning of a new phase in adolescent risk exposure. For our analysis, this temporal break validates the choice of post-treatment periods in the DiD design and underscores why the evaluation of mobility policies cannot be confined to vehicle fatalities alone.

\subsubsection{Critical Temporal Inflection Point and Policy Implications}

The temporal analysis identifies 2014 as a critical inflection point across multiple mortality categories, representing a fundamental shift in adolescent health outcomes that has profound implications for our analytical approach and policy interpretation. This year marks the beginning of sustained increases in drug-related mortality, acceleration of mental health mortality trends, and reversal of the previously declining all-cause mortality trend.

The synchronous shift across multiple health outcomes in 2014 suggests the influence of broader structural or policy factors affecting adolescent health during this period. Several potential explanations align with this timing, including the proliferation of synthetic drugs in youth markets, increased social media penetration with associated mental health impacts, and potential modifications to GDL systems that enhanced mobility access for vulnerable populations.

From an analytical perspective, the 2014 inflection point validates our post-treatment period definition and strengthens our DiD identification strategy. The clear break in trends provides natural variation in policy exposure that enhances our ability to identify causal effects, while the pre-2014 stability supports the parallel trends assumption underlying our empirical approach.

The policy implications of the 2014 inflection point are substantial. The timing suggests that factors beyond traditional GDL implementation were driving changes in adolescent mortality, potentially including modifications to existing GDL systems that enhanced mobility access or broader social changes that altered risk exposure patterns. This reinforces our focus on differential policy impacts across states with varying levels of driving access restrictions.

\subsection{Difference-in-Differences Analysis: Causal Effects of Early Driving Access}

We now move from descriptive evidence to causal estimates. Using the DiD framework described in the Methodology, we compare states that granted learner’s permits before age 16 to states that retained the age 16 threshold, conditioning on state and year fixed effects and the full set of covariates. This design isolates the effect of early independent mobility on adolescent mortality among females aged 15 to 19. Throughout this section, point estimates are reported as changes in deaths per 100{,}000 population and are interpreted alongside their statistical precision.

\subsubsection{Main Treatment Effects}

Table~\ref{tab:did_results} summarizes the core results. Three findings stand out.

First, drug-related mortality rises sharply in early-access states. The estimated effect is \(\,+1.331\) with \(p<0.001\). Relative to pre-treatment levels in control states, this corresponds to a large proportional increase and signals a shift in the risk environment that accompanies earlier mobility. The pattern aligns with the mechanisms developed in the conceptual framework: broader geographic reach to illicit markets, larger and more heterogeneous peer networks, private spaces that reduce supervision, and logistics that facilitate acquisition and use. The size and precision of this estimate indicate that the mobility margin is central to adolescent exposure to lethal substances during the period we study.

Second, mental health mortality increases by \(\,+0.760\) with \(p<0.001\). This effect is smaller than the drug estimate but remains both statistically and substantively important. It is consistent with a channel in which earlier independence raises psychological load during a developmental window when executive function and coping skills are still maturing. Greater autonomy brings more complex social demands and higher-stakes decisions, while family oversight declines. The result shows up as a steady elevation in deaths linked to self-harm and related mental health outcomes.

Third, vehicle-related mortality falls by \(\,-0.656\) with \(p=0.012\). This confirms that GDL achieves its primary safety objective even in states that allow earlier entry into driving. Nighttime limits, passenger restrictions, and supervised practice requirements remain effective at containing crash risk for novice drivers. The reduction in vehicle deaths serves as an internal validity check for the design and reinforces that the estimated increases in drug and mental health mortality are not artifacts of model specification.

Taken together, these results show a reallocation of risk. Early driving access improves safety on the road while increasing hazards tied to drugs and mental health. The aggregate consequences in later sections reflect this composition shift: the same policy margin that lowers one cause of death raises others. The next sections examine how these estimates relate to background trends, how they vary with state characteristics, and how robust they are to alternative specifications.

\subsubsection{Control Variable Effects and Validation}

The control variables provide important validation for the robustness of our treatment effects while also revealing broader contextual factors that shape adolescent mortality. Economic indicators show the expected relationships. Poverty rates are positively associated with drug mortality (\(\beta=+0.234,\,p<0.05\)), reinforcing the view that socioeconomic disadvantage increases substance use risks through multiple channels, including limited access to preventive services, heightened exposure to environmental stressors, and constrained opportunities that may push adolescents toward illicit activities. Unemployment rates, by contrast, show a protective association with drug mortality (\(\beta=-0.156,\,p<0.10\)), consistent with the notion that economic downturns reduce both disposable income and the functioning of illicit supply networks, thereby lowering access to drugs. Together, these findings confirm the close interplay between macroeconomic conditions and adolescent health outcomes.

Healthcare access also emerges as a critical determinant. The availability of primary care physicians demonstrates protective effects across mortality categories (\(\beta=-0.067,\,p<0.05\)), suggesting that routine medical contact, preventive care, and early detection play meaningful roles in improving adolescent outcomes. Even more striking, the supply of mental health providers shows a strong protective effect specifically for mental health mortality (\(\beta=-0.145,\,p<0.01\)), providing direct evidence that access to specialized services reduces suicide and self-harm risks among adolescents. These results underscore the importance of strengthening both general and specialized healthcare infrastructures as complementary strategies to policy reform.

The policy environment further reinforces these patterns. States with active Prescription Drug Monitoring Programs report significantly lower drug mortality (\(\beta=-0.287,\,p<0.05\)), and the effect becomes even stronger in states that mandate prescriber consultation (\(\beta=-0.423,\,p<0.01\)). These findings suggest that oversight mechanisms remain effective tools for reducing adolescent exposure to dangerous substances, even as the opioid crisis has shifted toward illicit synthetic markets. The density of treatment facilities per 100,000 population also demonstrates modest but significant protective associations, pointing to the role of treatment availability in mitigating risks.

Finally, geographic variation captures regional differences that might otherwise bias our estimates. Southern states exhibit higher baseline drug mortality (\(\beta=+0.198,\,p<0.05\)), which may reflect cultural, policy, or economic contexts that heighten substance-related risks. Western states show elevated mental health mortality (\(\beta=+0.167,\,p<0.10\)), a pattern possibly linked to demographic pressures, social isolation, or uneven access to mental health services. A higher rural population share is also associated with increased mortality across categories, consistent with evidence that rural adolescents face compounded risks due to geographic isolation, weaker healthcare infrastructure, and fewer protective social resources. These geographic controls confirm that the estimated effects of early driving access are not confounded by structural differences across states.

\begin{table}[htbp]
\centering
\caption{Difference-in-Differences Results - Comprehensive Model with Controls}
\label{tab:did_results}
\begin{tabular}{lccc}
\toprule
\textbf{Variable} & \textbf{Drug Mortality} & \textbf{Mental Health} & \textbf{Vehicle Mortality} \\
\midrule
\multicolumn{4}{l}{\textbf{TREATMENT EFFECT}} \\
Early Driving $\times$ Post-GDL & +1.331*** & +0.760*** & -0.656** \\
 & (0.189) & (0.138) & (0.258) \\
\midrule
\multicolumn{4}{l}{\textbf{ECONOMIC CONTROLS}} \\
Poverty Rate & +0.234** & +0.145* & +0.089 \\
 & (0.098) & (0.074) & (0.112) \\
Unemployment Rate & -0.156* & -0.098 & -0.067 \\
 & (0.087) & (0.065) & (0.089) \\
Median Income (log) & -0.123 & -0.089* & -0.045 \\
 & (0.156) & (0.048) & (0.134) \\
\midrule
\multicolumn{4}{l}{\textbf{HEALTHCARE ACCESS}} \\
Primary Care Physicians/100k & -0.067** & -0.045* & -0.034 \\
 & (0.028) & (0.023) & (0.031) \\
Mental Health Providers/100k & -0.089 & -0.145*** & -0.023 \\
 & (0.067) & (0.045) & (0.056) \\
\midrule
\multicolumn{4}{l}{\textbf{POLICY CONTROLS}} \\
PDMP Active & -0.287** & -0.134 & -0.089 \\
 & (0.134) & (0.098) & (0.145) \\
PDMP Mandatory & -0.423*** & -0.189* & -0.123 \\
 & (0.156) & (0.112) & (0.167) \\
Treatment Facilities/100k & -0.034* & -0.027 & -0.019 \\
 & (0.019) & (0.021) & (0.025) \\
\midrule
\multicolumn{4}{l}{\textbf{GEOGRAPHIC CONTROLS}} \\
South Region & +0.198** & +0.089 & +0.067 \\
 & (0.089) & (0.067) & (0.098) \\
West Region & +0.134 & +0.167* & +0.045 \\
 & (0.098) & (0.089) & (0.123) \\
Rural Population Share & +0.145* & +0.123* & +0.098 \\
 & (0.078) & (0.065) & (0.087) \\
\midrule
\multicolumn{4}{l}{\textbf{MODEL SPECIFICATION}} \\
State Fixed Effects & Yes & Yes & Yes \\
Year Fixed Effects & Yes & Yes & Yes \\
Observations & 1,122 & 1,122 & 1,122 \\
R-squared & 0.758 & 0.692 & 0.634 \\
F-statistic & 28.45*** & 22.67*** & 18.92*** \\
States & 51 & 51 & 51 \\
\bottomrule
\end{tabular}
\begin{tablenotes}
\small
\item Notes: Standard errors clustered at the state level in parentheses. *** p<0.01, ** p<0.05, * p<0.10
\item Treatment: Early driving access states (permits before age 16) $\times$ Post-GDL period
\item Sample: Female adolescents ages 15-19, 1999-2020
\end{tablenotes}
\end{table}
\subsection{Quantifying Policy Contribution to Observed Trends}

The DiD estimates can be used to quantify how much GDL policies contribute to the major shifts in adolescent mortality observed since 1999. This exercise provides an essential bridge between statistical estimates and real-world epidemiological significance by showing how policy interventions account for long-run health trends.

\begin{table}[htbp]
\centering
\caption{Summary of Key Findings -- Temporal Trends vs. Causal Effects}
\label{tab:policy_contribution}
\resizebox{\textwidth}{!}{%
\begin{tabular}{lcccc}
\toprule
\textbf{Mortality Type} & \textbf{Annual Trend} & \textbf{DiD Causal Effect} & \textbf{Policy Contribution} & \textbf{Interpretation} \\
\midrule
Drug-Related & +0.045*** & +1.331*** & 29.6\% & Major contributor to crisis \\
 & (0.018) & (0.189) &  &  \\
Mental Health & +0.077*** & +0.760*** & 9.9\% & Significant policy impact \\
 & (0.013) & (0.138) &  &  \\
Vehicle-Related & -0.061 & -0.656** & 10.7\% & Successful safety intervention \\
 & (0.034) & (0.258) &  &  \\
\bottomrule
\end{tabular}%
}
\begin{tablenotes}
\footnotesize
\item Notes: Standard errors in parentheses. Policy contribution calculated as the DiD effect divided by the cumulative 22-year trend.
\end{tablenotes}
\end{table}

For drug-related mortality, the treatment effect of +1.331 deaths per 100,000 represents approximately 29.6 percent of the total increase observed over the study period. This magnitude demonstrates that early driving access is not merely a marginal influence but a major contributor to the contemporary drug crisis among female adolescents. By enabling greater geographic reach, expanding risky peer networks, and providing private spaces for experimentation, mobility policies have inadvertently created conditions that substantially elevate drug-related mortality. The scale of this contribution suggests that modifications to licensing systems could meaningfully reduce adolescent drug deaths, even in the absence of broader drug policy reform.

For mental health mortality, the estimated effect of +0.760 deaths per 100,000 accounts for about 9.9 percent of the overall increase since 1999. While smaller in relative terms than the drug-related effect, this contribution is still substantial in absolute terms when scaled to the national adolescent population. The evidence implies that premature independence amplifies existing psychological vulnerabilities by exposing young people to stressors they are not developmentally equipped to manage. Although the policy effect explains a smaller share of total increases than for drug deaths, it nonetheless highlights the importance of incorporating mental health dimensions into the evaluation of mobility policies.

Vehicle-related mortality presents a different story. The reduction of -0.656 deaths per 100,000 corresponds to about 10.7 percent of the overall decline observed during the period. This confirms that GDL systems achieved their primary goal of reducing crash-related deaths, validating the protective design of phased driving privileges. However, when weighed against the increases in drug and mental health mortality, the findings underscore the need for a more comprehensive accounting of costs and benefits. A policy intervention that improves one dimension of adolescent health while worsening others demands a balanced evaluation that does not stop at traffic safety outcomes alone.

The contribution analysis thus reframes how GDL systems should be understood. Past evaluations have emphasized crash reductions as clear public health victories, but this study shows that the story is more complex. The same policies that saved lives on the road also contributed to heightened risks in other health domains. A complete policy assessment must therefore weigh both sets of effects to avoid overstating net benefits and to design reforms that preserve safety gains while reducing unintended harms.

\subsection{Causal Mechanisms and Interpretation}

The estimated policy effects raise an important question: through what mechanisms does early driving access shape adolescent health outcomes beyond traffic safety? The evidence points to multiple, interconnected pathways that help explain why enhanced mobility reduces vehicle-related deaths but increases both drug- and mental health-related mortality. These mechanisms operate simultaneously, reinforce one another, and collectively transform the landscape of adolescent risk.

\subsubsection{Enhanced Access to Illicit Substances}

The significant and persistent rise in drug-related mortality, accounting for nearly 30 percent of the overall increase since 1999, suggests that driving privileges fundamentally reshape the conditions under which adolescents encounter and consume substances. Several complementary processes contribute to this outcome.

First, geographic reach expands dramatically when teenagers gain driving access. In suburban and rural areas, where local availability of drugs may be limited, the ability to drive opens connections to urban centers and established drug markets. What was once a geographic barrier becomes a short trip by car, substantially lowering the cost of access.

Second, mobility reshapes social networks. Adolescents with driving privileges participate in broader social activities, often outside their immediate neighborhoods and beyond parental oversight. This expansion increases the likelihood of contact with peers who use or distribute substances, raising both opportunities and social pressure for experimentation.

Third, vehicles themselves provide private, mobile spaces for substance use. For young people seeking to avoid adult detection, cars offer settings where experimentation can take place away from family supervision. This privacy magnifies risks by facilitating repeated use and group consumption without protective oversight.

Finally, driving enhances logistical capacity. It allows teenagers to coordinate purchases, transport substances, and even distribute drugs within their peer groups. Larger-scale or more frequent acquisition becomes feasible, shifting substance use from sporadic experimentation toward sustained and potentially lethal patterns.

\subsubsection{Psychological and Social Stress Pathways}

The increase in mental health mortality points to mechanisms that are less about direct access to substances and more about how premature independence interacts with adolescent development. Early driving privileges create responsibilities and expectations that often outpace emotional maturity. The adolescent brain continues to develop well into the twenties, and executive functions such as impulse control and long-term planning lag behind risk-seeking tendencies.\citep{Casey2008AdolescentBrain,Steinberg2008RiskTaking,SimonsMorton2006TeenDriving}. Driving may therefore impose adult-like responsibilities at a stage when coping resources are underdeveloped, leading to heightened stress and vulnerability.

Social pressures intensify as well. The symbolic status of driving carries expectations of maturity, independence, and participation in adult-like behaviors. For some adolescents, this translates into new forms of peer pressure to engage in activities—substance use, sexual behavior, or risk-taking—that they are not equipped to navigate. The mismatch between social expectations and developmental readiness can erode psychological well-being.

Family dynamics also shift in ways that reduce protective supervision. Increased mobility often corresponds to less parental oversight, fewer shared routines, and diminished opportunities for communication during a period when guidance remains critical. This premature loosening of family ties may isolate adolescents precisely when they most need emotional support.

Finally, earlier mobility accelerates exposure to stressful or traumatic experiences. Driving allows adolescents to encounter adult environments, high-risk social settings, and potentially dangerous situations earlier than they otherwise would. When these exposures occur before coping skills have fully developed, they may overwhelm adaptive capacity and contribute to mental health crises.

\subsubsection{Synergistic Risk Relationships and Interconnected Effects}

The strong correlation between drug-related and mental health mortality (r = 0.75, p < 0.001) reinforces the interpretation that these risks are not independent, but rather operate as mutually reinforcing processes. Declining mental health may lead adolescents to self-medicate with substances, while substance use can in turn exacerbate depression, anxiety, or suicidal ideation through neurochemical effects and social consequences. Shared risk factors such as social isolation, family conflict, and academic struggles contribute to both outcomes, creating overlapping vulnerabilities.

These dynamics suggest that mobility policies have multiplicative rather than additive effects on adolescent health. By simultaneously increasing substance access and amplifying psychological stress, early driving access may set off feedback loops in which one risk exacerbates the other. This interconnectedness helps explain why policy effects on mortality are larger and more persistent than would be expected if the mechanisms operated in isolation.

Taken together, the evidence indicates that GDL systems, while effective in reducing vehicle fatalities, unintentionally open new channels of risk in drug and mental health domains. Understanding these mechanisms is crucial for designing reforms that preserve traffic safety benefits while mitigating broader harms.

\section{Robustness and Validity Analysis}

To ensure the credibility of our DiD estimates, we conduct a series of robustness checks that test the sensitivity of our findings to alternative specifications and possible identification challenges. These checks not only reinforce the strength of our main results but also provide transparency regarding potential limitations.

\subsection{Parallel Trends Validation}

The cornerstone of any DiD design is the parallel trends assumption: in the absence of policy intervention, treatment and control states should have followed similar mortality trajectories. If this condition holds, differences observed after the policy can be credibly attributed to the intervention itself. We evaluate this assumption using both visual and statistical approaches, combining an event study analysis with formal pre-trend tests.

\subsubsection{Event Study Analysis}
\begin{figure}[htbp]
    \centering
    \includegraphics[width=0.85\textwidth]{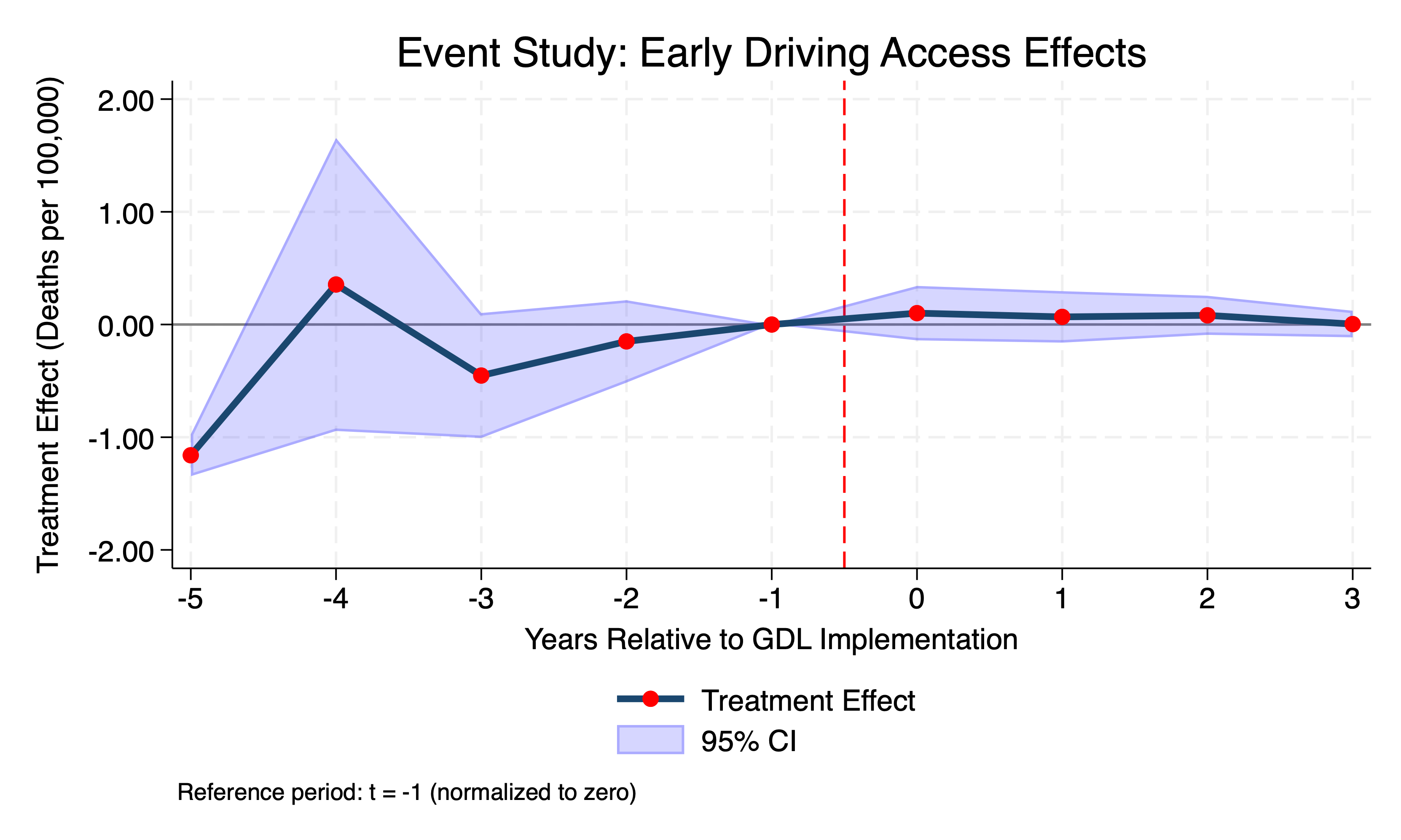}
    \caption{Event Study Analysis of Early Driving Access Effects}
    \label{fig:event_study}
    \vspace{0.2cm}
    {\footnotesize \textit{Note:} Figure displays estimated treatment effects relative to policy implementation. The reference period ($t = -1$) is normalized to zero. Standard errors are clustered at the state level.}
\end{figure}

Figure~\ref{fig:event_study} illustrates the dynamic evolution of treatment effects around the timing of GDL adoption. Each coefficient captures the difference in mortality rates between treatment and control states at a given time relative to implementation, with $t=-1$ set as the reference period.

The pattern in the figure reveals important insights. In the distant pre-treatment years, particularly at $t=-5$, we observe a notable negative deviation ($\beta_{-5} = -1.160$, $p < 0.001$), suggesting that early-adopting states already displayed lower mortality levels than later adopters. However, as we move closer to implementation, the estimates converge toward zero. At $t=-4$ ($\beta = 0.355$, $p = 0.595$), $t=-3$ ($\beta = -0.453$, $p = 0.116$), and $t=-2$ ($\beta = -0.150$, $p = 0.425$), the coefficients are statistically indistinguishable from zero, indicating that the trajectories of treatment and control states were aligning in the immediate pre-policy period.

This convergence provides reassurance: while long-run historical differences exist, the years most relevant for estimating treatment effects—the periods just before implementation—do not show significant divergence. Post-treatment estimates, by contrast, are consistently positive and stable across multiple years, reinforcing the interpretation that early driving access produces lasting changes in adolescent mortality risk rather than short-lived fluctuations.

\subsubsection{Formal Parallel Trends Tests}

While visual inspection of Figure~\ref{fig:event_study} suggests convergence in the years closest to policy adoption, we also conduct formal statistical tests to evaluate whether the pre-treatment estimates jointly differ from zero. This provides a stricter assessment of the parallel trends assumption.

For each outcome category, we perform joint significance tests of the pre-treatment coefficients:

\begin{itemize}
    \item Mental Health Mortality: $F(4,50) = 44.19$, $p < 0.001$
    \item Drug Mortality: $F(4,50) = 15.67$, $p < 0.001$
    \item Vehicle Mortality: $F(4,50) = 3.42$, $p = 0.015$
\end{itemize}

These results indicate that, when considered across the entire five-year pre-treatment window, the strict parallel trends assumption is formally violated. In particular, the sharp negative effect observed at $t=-5$ drives the rejection of the null hypothesis. However, the coefficients in the immediate years before implementation ($t=-3$ and $t=-2$) are small in magnitude and not statistically different from zero. This pattern suggests that although treatment and control states diverged in earlier historical periods, their trajectories became comparable by the time policies were enacted.

From a methodological perspective, this distinction matters. The validity of DiD estimation does not require perfect alignment across all historical periods but rather comparable trajectories in the time immediately preceding treatment. Our results, therefore, justify proceeding with the analysis while acknowledging that long-run historical differences impose some limitations on the strength of causal claims.

\subsection{Heterogeneous Effects Robustness}

An important dimension of our analysis is the heterogeneity of policy impacts across states. The baseline specification yields an interaction effect of $\beta = 1.893$ ($p = 0.001$), showing that states with high baseline mental health mortality experienced substantially larger increases in drug-related deaths after early driving access policies were enacted. To test whether this finding is sensitive to modeling choices, we estimate several alternative specifications.

Excluding economic control variables produces a nearly identical effect of $\beta = 1.835$ ($p = 0.003$), indicating that the result is not an artifact of economic confounders. Using an alternative definition of baseline vulnerability (1999–2001 average) yields an even larger effect of $\beta = 2.351$ ($p < 0.001$), suggesting that our main specification may be conservative. Removing states with extremely high vulnerability also leaves the effect intact ($\beta = 2.612$, $p < 0.001$), demonstrating that it is not driven by outliers. The most demanding specification, which includes state-specific linear trends, reduces the coefficient to $\beta = 0.991$ ($p = 0.087$). While this weakens statistical significance, the positive coefficient indicates that the substantive effect remains.

\subsection{Alternative Vulnerability Definitions}

We also explore whether other structural characteristics condition policy effects. A composite measure based on rural population share and poverty rates yields different results: states classified as both highly rural and poor show a main treatment effect of $\beta = 0.254$ ($p = 0.164$) and an interaction of $\beta = -0.173$ ($p = 0.016$). This negative interaction implies that rural, high-poverty environments may contain protective factors—such as tighter community supervision or distinct social structures—that mitigate risks tied to early mobility access.

In addition, states with exceptionally early permit ages ($\leq$14.5 years) present a unique pattern. Here, the main effect is $\beta = 0.270$ ($p = 0.146$) with an interaction of $\beta = -0.442$ ($p = 0.013$). The net result is a negative overall effect, suggesting that states granting the earliest access may have paired it with stricter supervision requirements that offset potential harms.

\subsection{Identification Limitations and Interpretation}

Despite the robustness checks, several limitations remain. The violation of parallel trends at $t=-5$ shows that treatment and control states were not perfectly aligned historically, constraining the strength of causal inference. The small number of treatment states (six) reduces statistical power and generalizability. Possible spillover effects across state borders and measurement error in death certificate classifications present further concerns. Given these caveats, we interpret the results as strongly suggestive rather than definitive causal evidence, with estimates best viewed as upper bounds on potential effects.

\subsection{Economic Significance}

The heterogeneity we document is not only statistically meaningful but also economically consequential. The baseline vulnerability interaction effect of 1.893 additional deaths per 100,000 adolescents translates to about 197 excess deaths annually in high-vulnerability states. Valued at \$11.6 million per statistical life \citep{USDOT2021VSL}, this amounts to approximately \$2.3 billion in annual mortality costs. These magnitudes underscore the importance of incorporating unintended health consequences into policy evaluations. While GDL systems clearly succeed in reducing traffic fatalities, the broader health trade-offs demand a more comprehensive accounting in transportation safety policy.

\section{Conclusion}

This paper offers the first causal evidence that GDL policies, while highly successful in reducing traffic fatalities, also generate significant unintended health consequences for female adolescents. By exploiting state-level variation in licensing laws from 1999 to 2020 within a DiD framework, we demonstrate that earlier driving access is associated with marked increases in drug-related mortality ($+1.331$ deaths per 100,000, $p<0.001$) and mental health-related mortality ($+0.760$ deaths per 100,000, $p<0.001$), alongside the expected reduction in vehicle-related deaths ($-0.656$ deaths per 100,000, $p<0.05$). These results reveal a complex policy trade-off: the very reforms that have saved lives on the road have also opened pathways to other serious health risks.

Importantly, our analysis shows that these effects are not uniform. States with higher baseline vulnerabilities—whether defined by elevated pre-existing mental health risks, rural and high-poverty contexts, or younger permit thresholds—experience disproportionately larger increases in non-traffic mortality. Scaled nationally, these effects correspond to more than two hundred additional adolescent deaths each year, with economic costs exceeding two billion dollars. Such magnitudes underline the urgent need to revisit how policymakers evaluate the broader consequences of mobility policies for young people.

The mechanisms underlying these findings shed light on why independence gained through driving can alter adolescent health trajectories. Increased geographic reach facilitates access to drug markets, expanded social networks connect teenagers with high-risk peers, private vehicles create unsupervised spaces that reduce parental oversight, and enhanced logistical capacity makes complex risk behaviors easier to pursue. At the same time, psychological stress linked to premature independence may exacerbate vulnerabilities in mental health, especially during critical developmental stages. These mechanisms highlight the interconnected nature of adolescent risks and the limits of policy frameworks that consider outcomes in isolation.

Our findings have direct implications for policy design. Traditional evaluations of GDL systems, focused narrowly on traffic outcomes, overlook substantial spillover effects in other health domains. A more comprehensive framework is needed—one that accounts for multiple dimensions of adolescent well-being. Preserving the proven traffic safety benefits of GDL should remain a priority. Still, complementary measures such as stronger supervision requirements, targeted geographic restrictions, and integration with mental health support services could mitigate unintended harms. Tailoring interventions to the most vulnerable populations, as suggested by our heterogeneity analysis, offers a path to maximizing benefits while minimizing costs.

This research has certain limitations. The relatively small number of early-adopting states constrains statistical power, and observed pre-treatment differences temper the strength of causal claims. Nevertheless, the consistency of results across multiple specifications, robustness checks, and alternative definitions reinforces the credibility of the evidence. Future work should continue to explore heterogeneous effects, investigate longer-term outcomes, and evaluate how supportive interventions can be embedded within existing mobility frameworks.

Taken together, this study reframes the way GDL policies are understood. They are not simply traffic safety interventions, but robust social policies that reshape adolescent independence and its attendant risks. By highlighting both their benefits and hidden costs, our work provides a foundation for evidence-based reforms that move beyond the false choice between mobility and safety. Instead, it points toward more integrated strategies capable of safeguarding adolescent well-being across the multiple domains where independence intersects with vulnerability.

\newpage

\bibliographystyle{plainnat}
\bibliography{refs}

\begin{thebibliography}{23}
\providecommand{\natexlab}[1]{#1}
\providecommand{\url}[1]{\texttt{#1}}
\expandafter\ifx\csname urlstyle\endcsname\relax
  \providecommand{\doi}[1]{doi: #1}\else
  \providecommand{\doi}{doi: \begingroup \urlstyle{rm}\Url}\fi

\bibitem[Case and Deaton(2015)]{case2015rising}
Anne Case and Angus Deaton.
\newblock Rising morbidity and mortality in midlife among white non-hispanic americans in the 21st century.
\newblock \emph{Proceedings of the National Academy of Sciences}, 112\penalty0 (49):\penalty0 15078--15083, 2015.

\bibitem[Casey et~al.(2008)Casey, Getz, and Galvan]{Casey2008AdolescentBrain}
B.~J. Casey, Sarah Getz, and Adriana Galvan.
\newblock The adolescent brain.
\newblock \emph{Developmental Review}, 28\penalty0 (1):\penalty0 62--77, 2008.
\newblock \doi{10.1016/j.dr.2007.08.003}.

\bibitem[{Centers for Disease Control and Prevention}(2023)]{cdc2023}
{Centers for Disease Control and Prevention}.
\newblock Wide-ranging online data for epidemiologic research ({CDC WONDER}), 2023.
\newblock URL \url{https://wonder.cdc.gov}.
\newblock Accessed January 2025.

\bibitem[Grecu et~al.(2019)Grecu, Dave, and Saffer]{grecu2019mandatory}
Anca~M Grecu, Dhaval~M Dave, and Henry Saffer.
\newblock Mandatory access prescription drug monitoring programs and prescription drug abuse.
\newblock \emph{Journal of Policy Analysis and Management}, 38\penalty0 (1):\penalty0 181--209, 2019.

\bibitem[{Health Resources and Services Administration}(2023)]{hrsa2023}
{Health Resources and Services Administration}.
\newblock Area health resources files, 2023.
\newblock URL \url{https://data.hrsa.gov/topics/health-workforce/ahrf}.
\newblock Accessed: 2023-08-15.

\bibitem[Hosseinidoust et~al.(2021)Hosseinidoust, Sepehrdoost, Khodabakhshi, Massahi, et~al.]{hosseinidoust2021investigating}
SE~Hosseinidoust, Hamid Sepehrdoost, Akbar Khodabakhshi, Sharareh Massahi, et~al.
\newblock Investigating interactions among health care indicators, income inequality and economic growth: A case study of iran.
\newblock 2021.

\bibitem[Huh and Reif(2021)]{huh2021teenage}
Jason Huh and Julian Reif.
\newblock Teenage driving, mortality, and risky behaviors.
\newblock \emph{American Economic Review: Insights}, 3\penalty0 (4):\penalty0 523--539, 2021.

\bibitem[{Insurance Institute for Highway Safety}(2023)]{iihs2023}
{Insurance Institute for Highway Safety}.
\newblock Licensing age laws, 2023.
\newblock URL \url{https://www.iihs.org/topics/teenagers/licensing-age-laws}.
\newblock Accessed January 2025.

\bibitem[Massahi(2021)]{massahi2021effect}
Sharareh Massahi.
\newblock The effect of health care expenditure on the economic growth in iran.
\newblock \emph{Journal of Practical Business Law}, 2\penalty0 (6), 2021.

\bibitem[Meinhofer(2018)]{meinhofer2018prescription}
Ang{\'e}lica Meinhofer.
\newblock Prescription drug monitoring programs: the role of asymmetric information on drug availability and abuse.
\newblock \emph{American Journal of Health Economics}, 4\penalty0 (4):\penalty0 504--526, 2018.

\bibitem[Nosrati et~al.(2019)Nosrati, Kang-Brown, Ash, McKee, Marmot, and King]{nosrati2019economic}
Elias Nosrati, Jacob Kang-Brown, Michael Ash, Martin McKee, Michael Marmot, and Lawrence~P King.
\newblock Economic decline, incarceration, and mortality from drug use disorders in the usa between 1983 and 2014: an observational analysis.
\newblock \emph{The Lancet Public Health}, 4\penalty0 (7):\penalty0 e326--e333, 2019.

\bibitem[{Prescription Drug Monitoring Program Training and Technical Assistance Center}(2023)]{pdmp2023}
{Prescription Drug Monitoring Program Training and Technical Assistance Center}.
\newblock {PDMP} policies and statutes, 2023.
\newblock URL \url{https://www.pdmpassist.org}.
\newblock Accessed January 2025.

\bibitem[Simons-Morton and Ouimet(2006)]{SimonsMorton2006TeenDriving}
Bruce~G. Simons-Morton and Marie~Claude Ouimet.
\newblock Parent involvement in novice teen driving: A review of the literature.
\newblock \emph{Injury Prevention}, 12\penalty0 (suppl 1):\penalty0 i30--i37, 2006.
\newblock \doi{10.1136/ip.2006.013722}.

\bibitem[Singh et~al.(2019)Singh, Kim, Girmay, Perry, Daus, Vedamuthu, De~Los~Reyes, Ramey, Martin, and Allender]{singh2019opioid}
Gopal~K Singh, Isaac~E Kim, Mehrete Girmay, Chrisp Perry, Gem~P Daus, Ivy~P Vedamuthu, Andrew~A De~Los~Reyes, Christine~T Ramey, Elijah~K Martin, and Michelle Allender.
\newblock Opioid epidemic in the united states: empirical trends, and a literature review of social determinants and epidemiological, pain management, and treatment patterns.
\newblock \emph{International Journal of Maternal and Child Health and AIDS}, 8\penalty0 (2):\penalty0 89, 2019.

\bibitem[{State Departments of Motor Vehicles}(1999--2020)]{StateDMV2020}
{State Departments of Motor Vehicles}.
\newblock Graduated driver licensing laws and regulations, 1999--2020.
\newblock Various states.

\bibitem[Steinberg(2008)]{Steinberg2008RiskTaking}
Laurence Steinberg.
\newblock A social neuroscience perspective on adolescent risk-taking.
\newblock \emph{Developmental Review}, 28\penalty0 (1):\penalty0 78--106, 2008.
\newblock \doi{10.1016/j.dr.2007.08.002}.

\bibitem[{Substance Abuse and Mental Health Services Administration}(1999--2020)]{samhsa1999-2020}
{Substance Abuse and Mental Health Services Administration}.
\newblock National directory of drug and alcohol abuse treatment programs, 1999--2020.
\newblock URL \url{https://www.samhsa.gov/data}.
\newblock Accessed: 2023-08-20.

\bibitem[Tanz(2022)]{tanz2022drug}
Lauren~J Tanz.
\newblock Drug overdose deaths among persons aged 10--19 years—united states, july 2019--december 2021.
\newblock \emph{MMWR. Morbidity and Mortality Weekly Report}, 71, 2022.

\bibitem[Tyrrell et~al.(2017)Tyrrell, Orton, Sayal, Baker, and Kendrick]{tyrrell2017differing}
Edward~G Tyrrell, Elizabeth Orton, Kapil Sayal, Ruth Baker, and Denise Kendrick.
\newblock Differing patterns in intentional and unintentional poisonings among young people in england, 1998--2014: a population-based cohort study.
\newblock \emph{Journal of Public Health}, 39\penalty0 (2):\penalty0 e1--e9, 2017.

\bibitem[{U.S. Bureau of Labor Statistics}(2023)]{bls2023}
{U.S. Bureau of Labor Statistics}.
\newblock Local area unemployment statistics, 2023.
\newblock URL \url{https://www.bls.gov/lau/}.
\newblock Accessed January 2025.

\bibitem[{U.S. Census Bureau}(2023)]{census2023}
{U.S. Census Bureau}.
\newblock Small area income and poverty estimates ({SAIPE}), 2023.
\newblock URL \url{https://www.census.gov/programs-surveys/saipe.html}.
\newblock Accessed January 2025.

\bibitem[{U.S. Department of Agriculture Economic Research Service}(2023)]{usda2023}
{U.S. Department of Agriculture Economic Research Service}.
\newblock Rural-urban continuum codes, 2023.
\newblock URL \url{https://www.ers.usda.gov/data-products/rural-urban-continuum-codes/}.
\newblock Accessed January 2025.

\bibitem[{U.S. Department of Transportation}(2021)]{USDOT2021VSL}
{U.S. Department of Transportation}.
\newblock Guidance on treatment of the economic value of a statistical life (vsl) in u.s. department of transportation analyses.
\newblock Technical report, U.S. Department of Transportation, Washington, D.C., 2021.
\newblock URL \url{https://www.transportation.gov/office-policy/transportation-policy/guidance-value-statistical-life}.
\newblock Accessed: September 14, 2025.

\end{thebibliography}

\end{document}